\documentclass[twocolumn,twoside,slac_two]{revtex4}
\usepackage{graphicx}
\usepackage{fancyhdr}
\usepackage{graphics}
\usepackage{epstopdf}
\usepackage{textpos}
\begin{document}

\title{\centering NEUTRINO INTERACTIONS}
\author{
\centering
\includegraphics[scale=0.15]{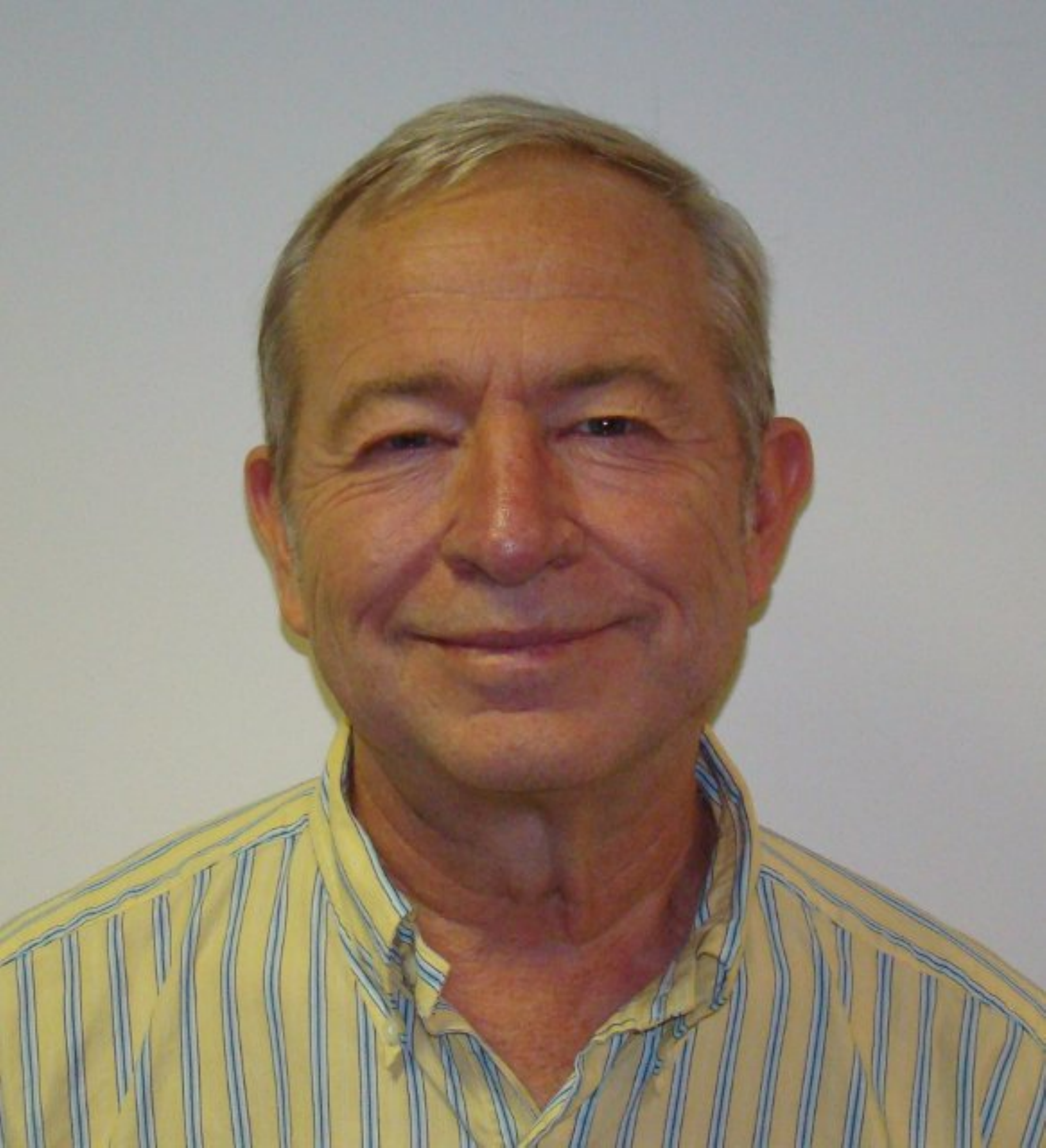} \\
\begin{center}
R.D. Ransome
\end{center}}
\affiliation{\centering Rutgers The State University of New Jersey, 
Piscataway, NJ, 08854, USA}
\begin{abstract}

The advent of very high intensity neutrino beams for the study of neutrino 
oscillations has also made possible a new generation of experiments which will 
study neutrino interactions on different nuclei with
unprecedented precision. The use of the neutrino as a probe of nucleon structure 
provides unique information not available with other probes.  In addition, an improved 
understanding of the neutrino interaction cross section and the resulting final states 
will reduce systematic uncertainties in neutrino oscillation experiments.
  A review of recent interaction measurements
 for neutrino energies
in the energy range of 500 MeV to a few GeV will be given, with an
emphasis on quasi-elastic scattering.

\end{abstract}

\maketitle
\thispagestyle{fancy}


\section{INTRODUCTION}

The discovery of neutrino oscillations has opened a new era of neutrino physics.
Very intense neutrino beam lines have been built at CERN, JPARC, 
 and Fermilab in
the past decade with the goal of studying neutrino oscillations in detail. These
cover a wide range of energies, from less than 1 GeV for the MiniBooNE, 
SciBooNE (Fermilab)\cite{boone}, and
T2K (JPARC)\cite{jparc} experiments, to peak energies of a few GeV with long tails up to greater than
50 GeV for the NuMI (Fermilab)\cite{numi} and CNGS (CERN)\cite{cngs} beam lines.
Accurate interpretation of oscillation experiments, and a thorough understanding of the systematic
uncertainties of the measurements, requires knowledge of both inclusive cross sections 
and details of final state characteristics.  Experiments have been designed to 
improve those measurements.  However, neutrino 
beams are so intense that they have also opened possibilities for using the neutrino
as a scattering probe to explore nucleon structure and to study the interaction of
neutrinos with nuclei with unprecedented detail.  It is this aspect of neutrino interactions
that I want to concentrate on in this report.

The neutrino has many attractive aspects as a scattering probe.  It interacts only through the
well understood weak interaction which makes it an ideal probe of the weak structure of the
nucleon.  It has a unique flavor sensitivity and, by combining neutrino and antineutrino scattering,
provides flavor separation not available by any other means.  Unfortunately it has a number
of unattractive aspects.  The incoming neutrino cannot be tracked, nor can the outgoing neutrino
for neutral current interactions be tracked.  Even with the intense beams currently available, massive targets (typically
many tons) are needed to get adequate rates.  The incident neutrino energy for any particular
interaction is not known, and beams have a broad range of energies for on axis beams, and even
off axis beams have significant (over 100 MeV) widths.  Finally, interpretation of any results
require detailed Monte Carlo simulations, and the validity of those simulations depends on the
accuracy of the input data.  

The fact that the incident neutrino energy is not known for any particular interaction is of
importance for oscillation experiments because the critical oscillation parameter is L/E (distance
over energy).  The energy is inferred from the energy and angle of the final state muon (for charged
current interactions), along with observed energy of the other final state particles.  Because the
interaction nucleus is typically carbon or heavier nuclei, energy can be unobserved because
of binding energy, final state neutrons, or other particles not well measured.  This aspect is
what makes understanding the final state characteristics so important to oscillation experiments.

Prior to the recent series of experiments, most cross section data were taken
with bubble chambers.  Very little data existed for at lower energies, especially for
heavier targets.  A summary of neutrino total cross sections is shown in Fig. 1.  As is clearly
evident, statistics are poor for neutrino energies of less than a few GeV.  

\begin{figure}[h]
 \includegraphics[width=60mm]{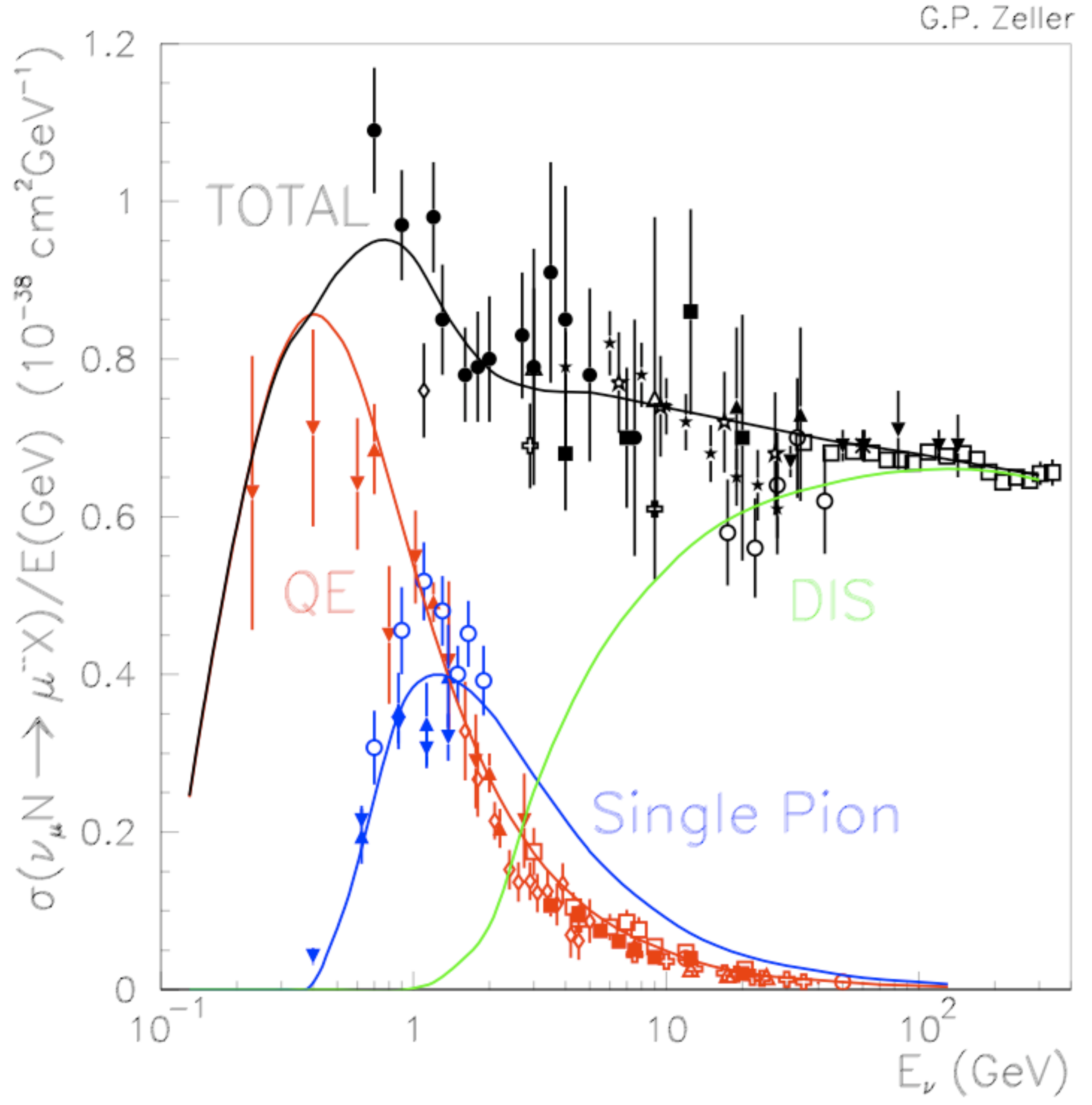}%
 \caption{\label{csec-all} Summary of cross sections measurements prior to 2006.  Plot courtesy of 
G. Zeller\cite{Zeller} }
 \end{figure}

A rather complete summary of neutrino interactions was given at PIC2010\cite{Wilk2010}, including a
discussion of higher energy measurements.  In this talk I will concentrate
on recent measurements of the quasi-elastic cross section at lower energies, and some data on coherent
scattering.

\section{QUASI-ELASTIC SCATTERING}

\subsection{Background}

Quasi-elastic scattering refers to the charged current interaction 
$\nu n \rightarrow \mu^{-}p$ or $\bar{\nu} p \rightarrow \mu^{+} n$.
The reaction is of interest for several reasons.  It is important for 
oscillation experiments because it has the best measure of the neutrino
energy.  For a true two body interaction, the neutrino energy can
be determined from the muon kinematics alone.  If the interaction occurs
in a nucleus, fermi momentum, the binding energy and final state interactions
smear and shift the estimated energy, but it still remains the best
measure of the incident neutrino energy.  For higher neutrino energies, above a few GeV,
the quasi-elastic cross section is nearly constant and can thus serve as a measure of the
neutrino flux.  For experiments using neutrino energies below one GeV, such
as MiniBooNE and T2K, the quasi-elastic interaction is the dominant part of
the cross section.

The reaction is of intrinsic interest for studying the nucleon because it is
the best way to determine the weak form factor of the nucleon.  The neutrino-nucleon
cross section can be written as a function of the vector form factors and the
axial form factor.  The vector form factors are taken from electromagnetic scattering:

$$\frac{d\sigma}{dQ^{2}}= \frac{M^{2}G_{F}^{2}cos^{2}(\theta_{c})}{8\pi E_{\nu}^{2}}[A-B(s-u)+C(s-u)^{2}]$$

\noindent where A, B, and C are functions of the electromagnetic and axial form factors,
$s$ and $u$ are the usual Mandelstam variables, $M$ is the nucleon mass,
$G_F$ is the Fermi constant, $\theta_{c}$ is the Cabbibo angle, and $E_\nu$ is
the neutrino energy.  Except for the neutron electric form factor, which is small
and has a different functional form, the nucleon form factors are approximately
described by the same functional form:

$$G_{E,M,A}(Q^{2})= \frac {G(0)}{(1+Q^{2}/M_{V,A}^2)^2}$$

\noindent where E, M, and A designate the electric, magnetic, and axial form factors,
and $M_{V,A}$ is a parameter called the vector or axial mass.  Although, the dipole form for
the form factors is empirical, and there is no particular physical significance to
the vector or axial mass, the form does have the correct behavior expected from 
perturbative QCD for very large $Q^2$.

\subsection{Axial Form Factor}

In the functional form given above, the vector form factors are determined from electron scattering and
the $Q^{2}=0$ value for $G_A$ is determined from neutron beta decay, leaving only one free parameter
to fit the $Q^2$ distribution for neutrino scattering: the axial mass $M_A$.  Prior to the 1990's, 
most determinations of the axial mass were done using neutrino scattering on deuterium or hydrogen targets.
Although the statistical uncertainties were typically 10-20\%, the experiments were in generally good
agreement with a world average for $M_A$ of about 1.03.  

\begin{figure}[h]
 \includegraphics[width=65mm]{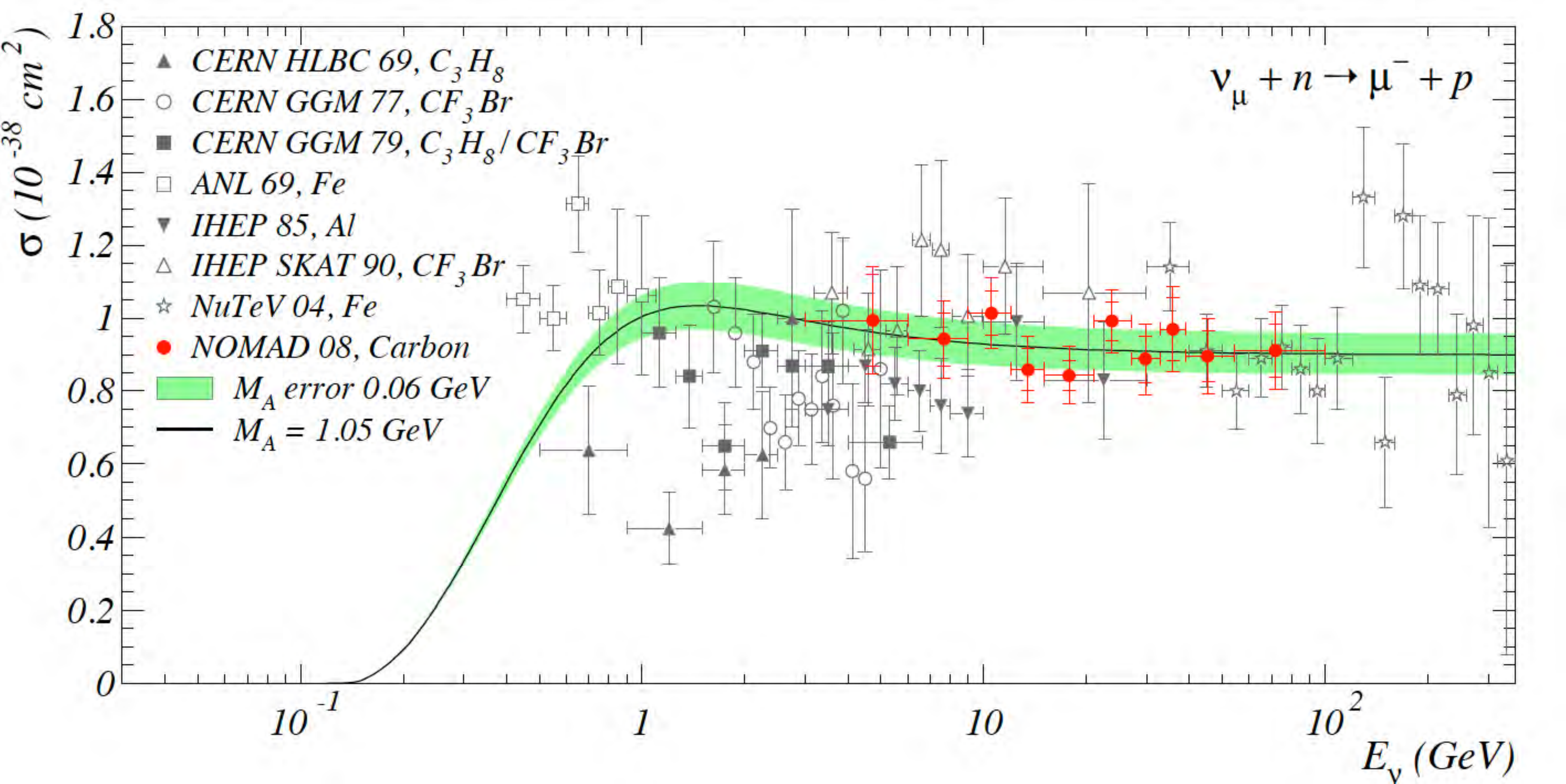}%
 \caption{Axial mass as determined in the NOMAD experiment, along with previous data.}
\label{nomad}
 \end{figure}

In the 1990's one of a new generation of oscillation experiments, NOMAD\cite{NOMAD}, measured the 
axial form factor as part of its studies\cite{Lyub2009} for neutrino and antineutrino
scattering with energies between 5 and 100 GeV.  The tracking detectors also served as the
scattering target and consisted of a mixture of elements, but were primarily carbon.  
As can be seen in Fig.~\ref{nomad}, the quasi-elastic cross section
and extracted axial mass were in generally good agreement with previous measurements.

The next generation of oscillation experiments took place at Fermilab in the Booster Neutrino Beamline,
with neutrino energies between about 0.5 and 1 GeV.  The first experiment, MiniBooNE, was
designed to check the LSND experiment. LSND observed oscillation values which,
combined with other oscillation results, would imply sterile neutrinos.  
The MiniBooNE detector consisted of a 6.1 m diamter tank of mineral oil (CH$_2$) viewed
by 1280 eight-inch phototubes.  Because intrinsic impurities present in
 mineral oil scintillate,
both Cerenkov light and scintillation light were detected.

A second detector, SciBooNE, was added on the same beam line later.  The SciBooNE detector
was orignally used at KEK (called SciBar). The detector consists of 14,436 strips of 
plastic scintillator (CH), each 1.3 cm x 2.5 cm x 300 cm, with wavelength shifting fiber readout.
It was moved to the Booster Neutrino Beamline in 
2007 and an electromagnetic calorimeter and muon range stack were added.  

\begin{figure}[h]
 \includegraphics[width=70mm]{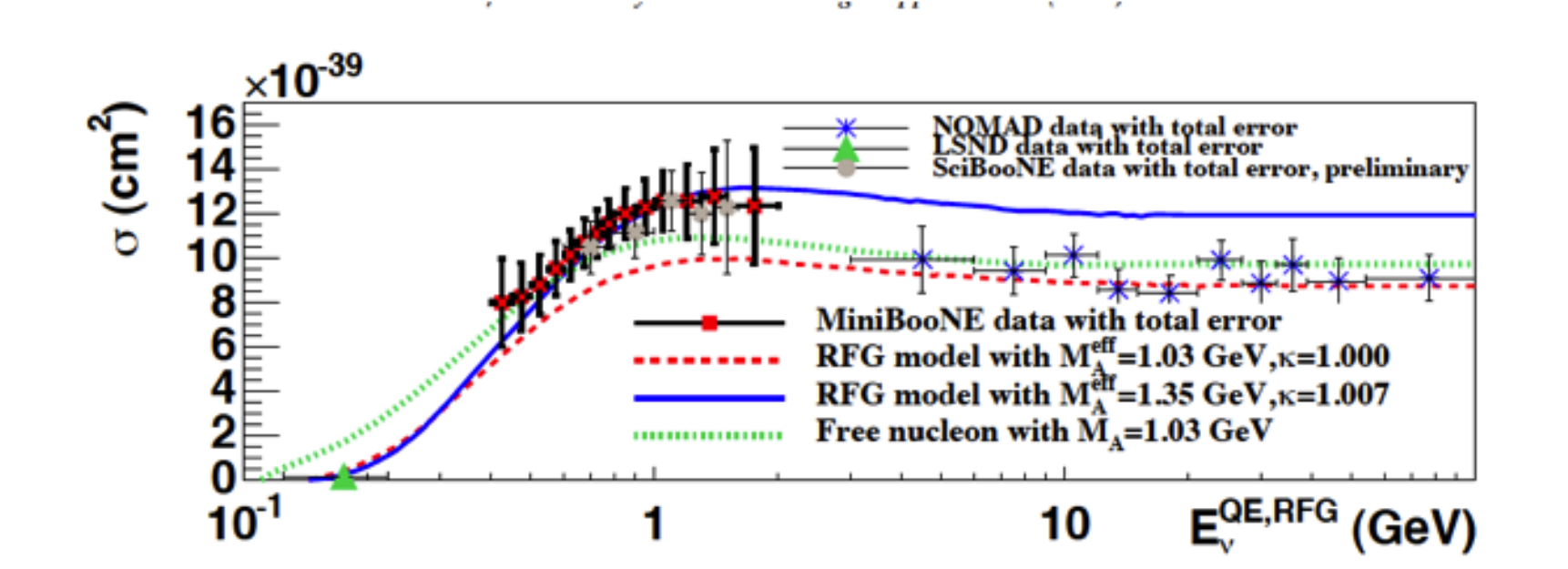}%
 \caption{Axial mass as determined by MiniBooNE and SciBooNE, along with the NOMAD data.}
\label{minib}
 \end{figure}

\begin{figure}[h]
 \includegraphics[width=70mm]{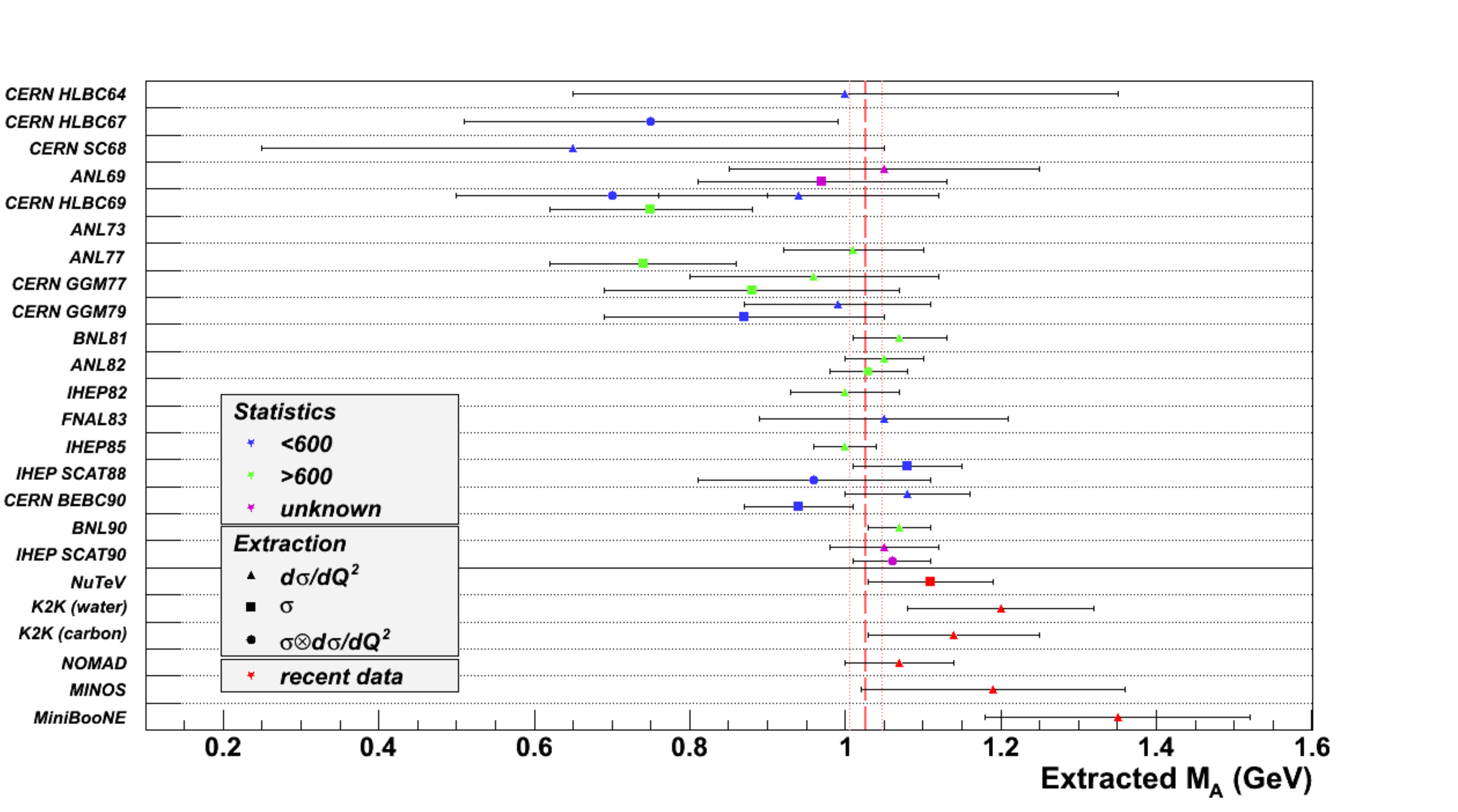}%
 \caption{Experimental measurements of the axial mass\cite{ben}.  }
\label{ma-zeimer}
 \end{figure}

Both MiniBooNE and SciBooNE were able to extract the axial mass in the lower energy range,
and both primarily from carbon\cite{miniboone,sciboone}.  As seen in Fig.~\ref{minib}, the
MiniBooNE and SciBooNE data are in good agreement with each other.  However, the best
fit to the data required a substantially higher value of $M_A$, of about 1.35 GeV.  The
gives a significant discrepancy with the higher energy data.  The form factor cannot
depend on the incident neutrino energy, so the difference must be due to some other effect.
Other more recent experiments have also found values of $M_A$ that are higher than
previous measurements, as shown in Fig.~\ref{ma-zeimer}.  The one common characteristic of
the later experiments with higher values of $M_A$ is that they were done at lower energy
and with heavier nuclear targets (i.e. not deuterium).  A possible solution of this
discrepancy may reside in the issues with the electromagnetic form factors or nuclear 
effects at low energy.

\subsection{Nuclear effects and the electromagnetic form factors}

\begin{figure}[h]
 \includegraphics[width=60mm]{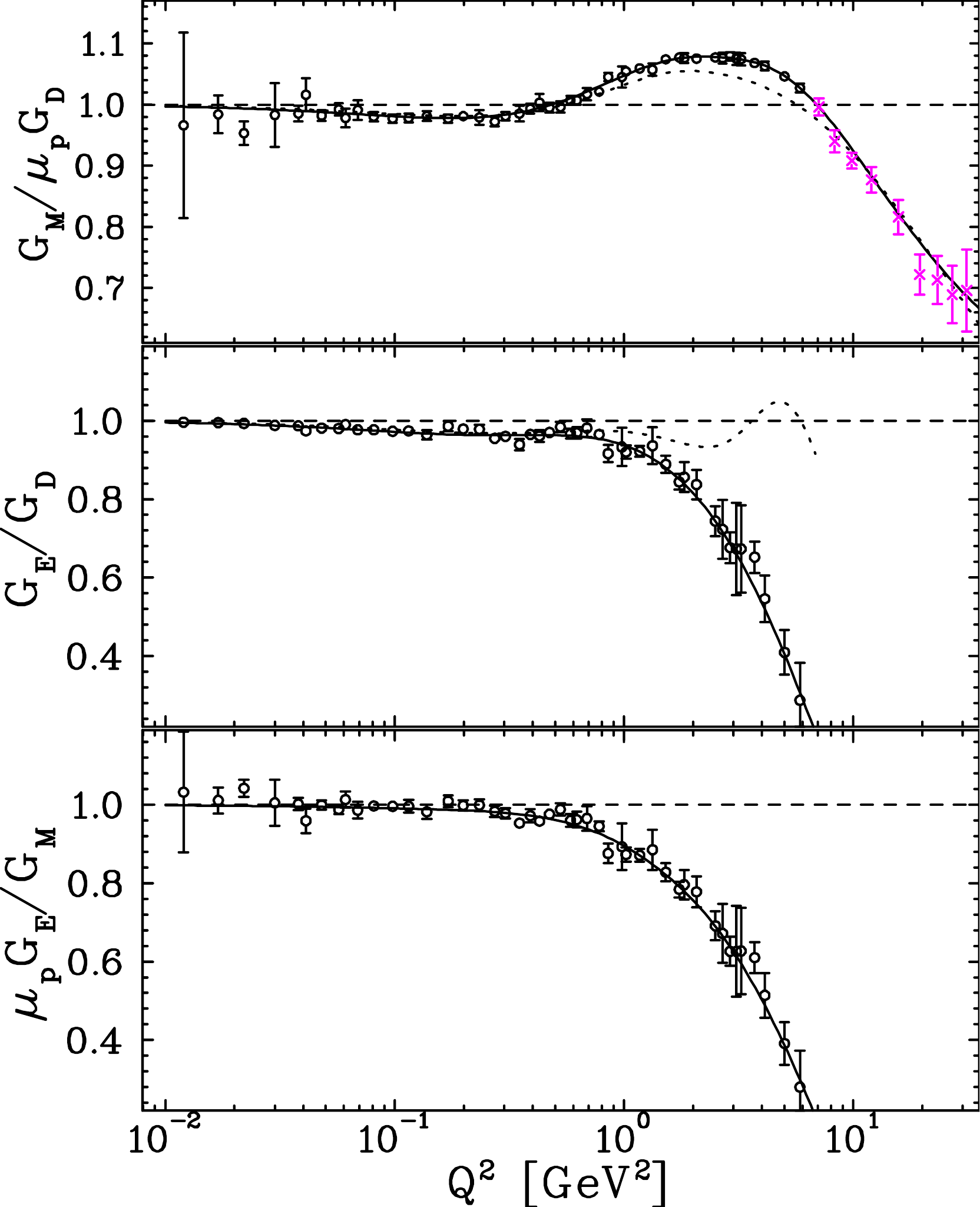}%
 \caption{Proton electromagnetic form factors from Ref.~\cite{arrington}.}
\label{gegmfig}
 \end{figure}

The history of the determination of the electromagnetic form factors
provides some interesting lessons for the determination of the axial form factor.
Early measurements showed good agreement with the dipole form of the form factor,
with a vector mass of 0.84 GeV.  Both the electric and magnetic form factors of
the proton appeared to have the same form, and the ratio appeared to be nearly
constant with $Q^2$.  However, beginning in the late 1990's, a series of experiments
using polarization transfer to determine the form factor ratio
showed deviations from the dipole form even at low $Q^2$, and a substantial change in
the ratio of the electric to magnetic form factors as $Q^2$ increased\cite{gegm}.  
Figure~\ref{gegmfig} shows a global fit to the world's data.  The important point
to be noticed here is that as $Q^2$ gets larger than 1 GeV$^2$, both $G_E$ and $G_M$
have significant deviations (about 10\% by $Q^2$ of 2 GeV$^2$) from the dipole form.
As measurements of the axial form factor become more precise, especially at higher
$Q^2$, potential deviations from the dipole form will have to be taken into account.

\begin{figure}[h]
 \includegraphics[width=60mm]{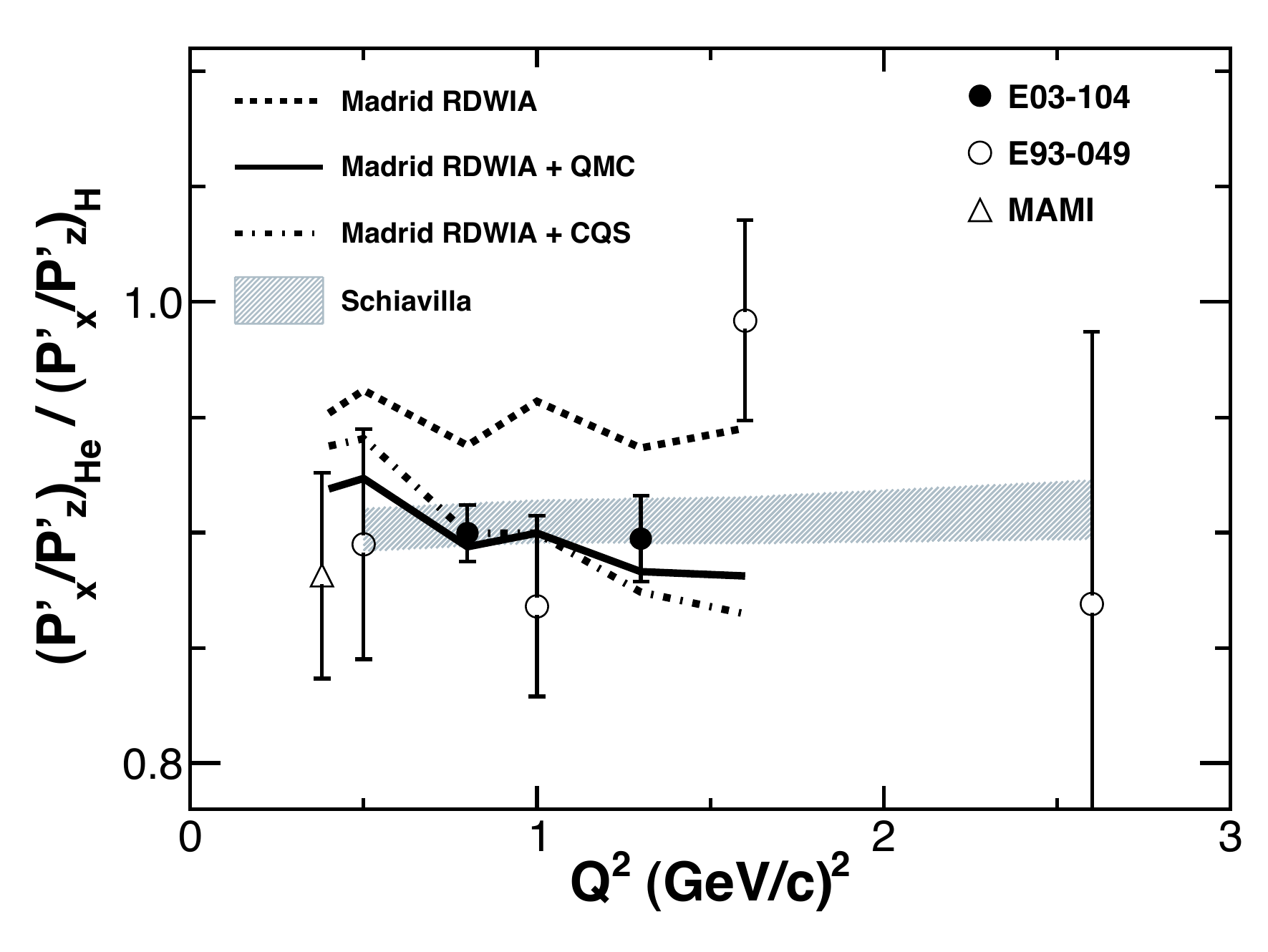}%
 \caption{Ratio of polarization transfer for the $^4$He$(\vec{e},e'\vec{p})^{3}$H reaction
compared to $^1$H$(\vec{e},e'\vec{p})$\cite{paolone}.  The ratio is sensitive to
medium modifications of the proton form factor.}
\label{gegmhe}
 \end{figure}

The electromagnetic form factors have been determined from electron scattering from
hydrogen targets.  Similar measurements were also made using the polarization
technique for the reaction $^4$He$(\vec{e},e'\vec{p})^{3}$H\cite{paolone}.  As is 
shown in Fig.~\ref{gegmhe}, the ratio differs from one by several percent.  Much of
the difference is explained from conventional nuclear effects.  However, conventional
models plus a modification of the nucleon form in the nuclear medium as predicted
by Saito {\it et al.}\cite{saito} gives a better fit to the data.  

The model from Ref.~\cite{saito} also predicts modification of the axial form factor, at the
level of 10\% for $Q^2$ less than 1 GeV$^2$.  However, the model predicts a {\it smaller}
value of $G_A$ in the medium, which is opposite of the apparent effect observed by
MiniBooNE/SciBooNE.  The question of a medium modification of the EM and axial
form factors is still open, and needs to be taken into account when extracting form
factors from the nucleons in the nuclear medium.

\begin{figure}[h]
 \includegraphics[width=60mm]{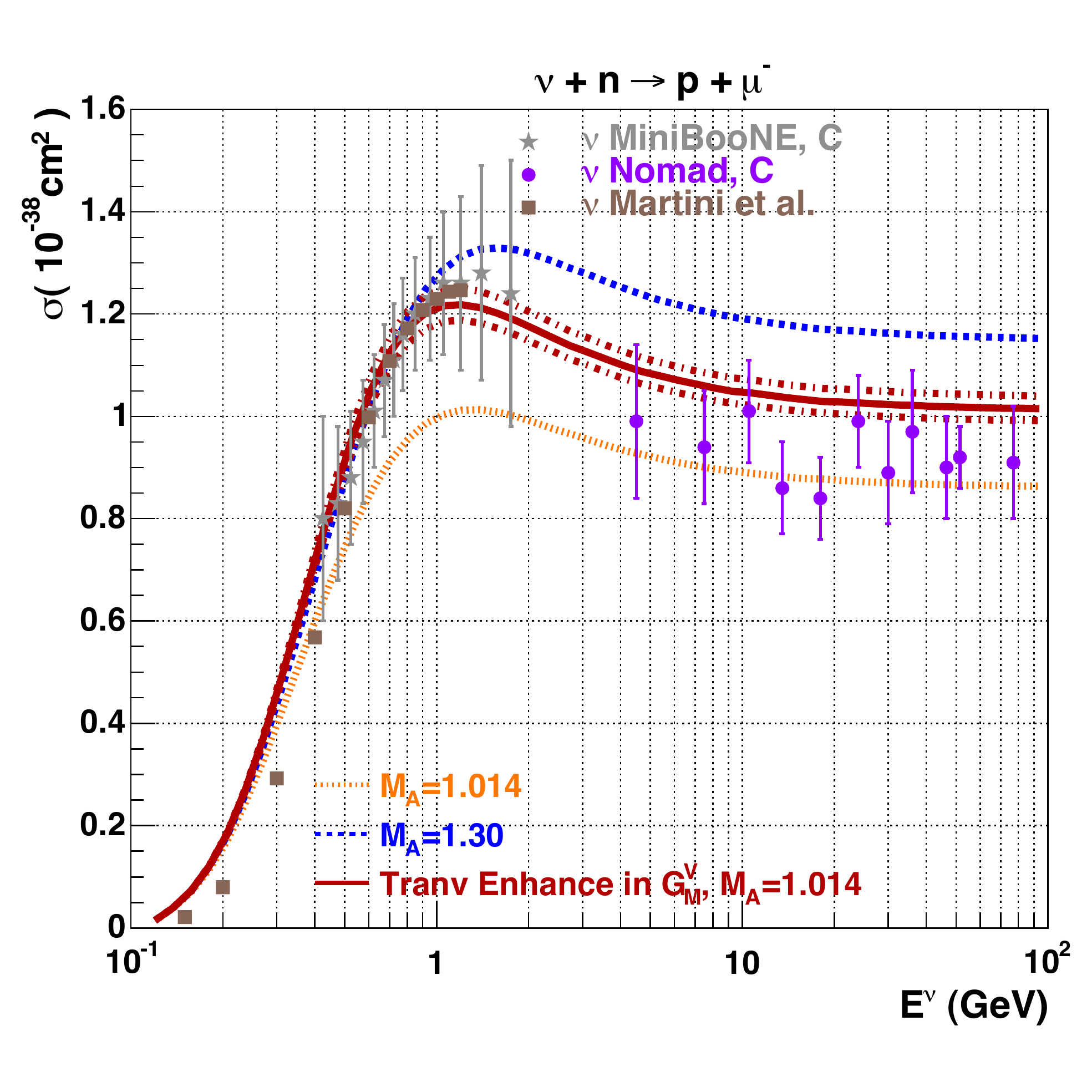}%
 \caption{Cross section predictions of the transverse enhancement model of Bodek {\it et al.}\cite{bodek}.
The model predicts the increased cross section for neutrino energies near 1 GeV compared to scattering
from deuterium, as well as the decrease as the neutrino energy increases.}
\label{bodekf}
 \end{figure}

Finally, another nuclear effect, which has been confirmed over the last decade, is the
enhancement of the transverse nuclear response compared to the free nucleon in
inclusive electron scattering\cite{carlson}.  The enhancement is attributed to 
meson exchange currents and short range correlations,
and is present even in $^4$He, at about the same level as in
heavier nuclei.  Bodek {\it et al.}\cite{bodek} have shown that the enhancement can be
modeled as an effective change of the nucleon magnetic form factor in nuclei.
The effect is most pronounced at low $Q^2$ and causes an increase in the cross section
compared to expectations from using the free form factors.  As the neutrino energy
increases, the maximum $Q^2$ increases, and the value of $M_A$ needed to fit the data
moves toward the free values.  As shown in Fig.~\ref{bodekf}, the model provides a
good description of both the low and high energy data, and appears to resolve the 
inconsistency between the two.  

To conclude this section, we see that the most recent experiments, with their much
improved statistics, have brought new interest to an old topic. Neutrino scattering
experiments have reached the stage of needing good models of nuclear effects as well
as potentially giving new information on the nuclear medium effects on the structure
of the nucleon.  Proper accounting for nuclear effects is a challenging theoretical
problem and even for the simpler case of electron scattering is not fully resolved.
High statistic neutrino scattering will no doubt provide even more challenges to theory.

\section{COHERENT PION PRODUCTION}

The second experimental result which has become of considerable interest is
coherent pion production.  Coherent pion production refers the reaction
$\nu + A \rightarrow \nu + \pi^{0} + A$ (neutral current production) and
$\nu + A \rightarrow \mu^{-}\pi^{+} + A$ (charged current production).  In each
case, the neutrino interacts with the entire nucleus, leaving it in the ground
state, with a single pion produced.  The kinematics of these interactions
give very little recoil energy to the nucleus, and the pion production is
forward peaked.  As with the quasi-elastic scattering,
this reaction has a particular interest for oscillation experiments, as well
as being an intrinsically interesting reaction.

Coherent pion production is a potentially large background for oscillation
experiments.  Neutral current production in particular is of a concern for 
experiments searching for electron neutrino appearance (such as NOvA\cite{nova}) because
the photon showers from a $\pi^0$ decay can easily be confused with an electron
shower if one photon either escapes detection or the two are too close to be distinguished.

\begin{figure}[ht]
 \includegraphics[width=60mm]{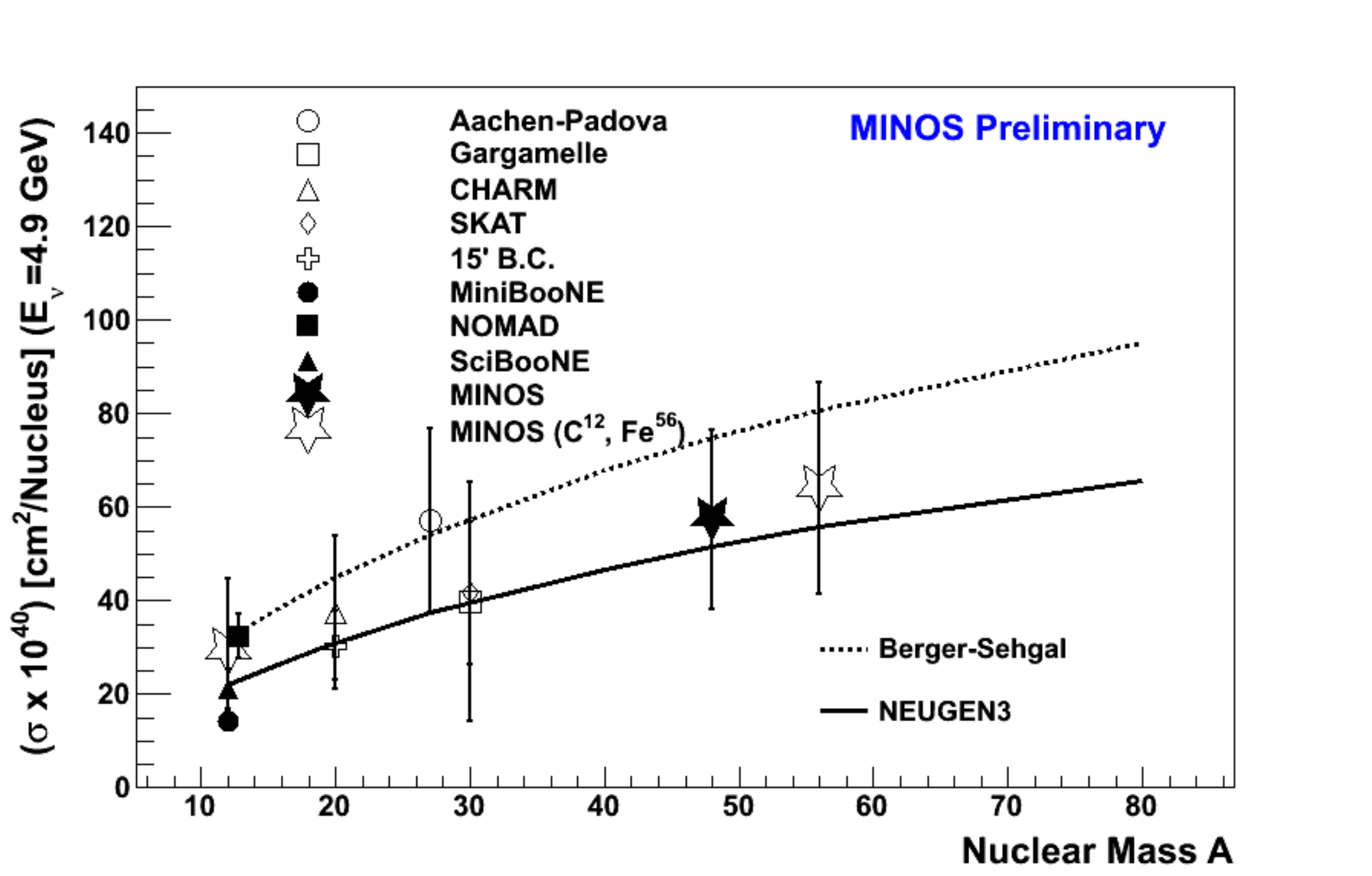}%
 \caption{MINOS coherent production cross section.}
\label{minos-coh}
 \end{figure}

A number of models\cite{models} exist which predict the coherent pion production cross section,
and the charged current to neutral current ratio, although the most widely used
is the Rein-Sehgal (RS) model\cite{rein}.  Recent reviews of the theoretical
status and comparison of models can be found in Refs.~\cite{alvarez} and \cite{boyd}.
Coherent pion production has been measured at higher energies by the MINOS experiment
at Fermilab\cite{MINOS,MINOS-coh}, and those data are in reasonable agreement with
RS, as shown in Fig.~\ref{minos-coh}. 

K2K\cite{k2k-coh} set  the first limit on CC coherent production near 1.2 GeV.
They found no evidence for this channel, with an upper limit of $0.6 \times 10^{-2}$ 
for the ratio of CC coherent to total CC interactions, below the prediction of RS
of about $1 \times 10^{-2}$.  SciBooNE\cite{scib-coh} has
observed NC coherent production at the level of $1.1 \times 10^{-2}$, in approximate
agreement with RS, but they also found CC coherent to be suppressed, with
a CC/NC ratio of only $0.14^{+0.30}_{-0.28}$.  RS and other recent models\cite{Hern} predict a
ratio of CC/NC to be 1 to 2.  This rather large discrepancy between models and data remains
a mystery.

\section{EXPERIMENTS}

As we have seen above, the most recent experiments have given resuls which appear
inconsistent with previous experiments and have placed new demands on theoretical explanations,
in particular on nuclear dependence of effects.  There is a need for more
precise measurements and studies of nuclear effects, especially at energies
below a few GeV.  In this section I want to discuss two new experiments which
significant potential to improve the experimental situation: MINERvA, at Fermilab,
and T2K at JPARC.

\subsection{MINERvA}

MINERvA\cite{min-coll,panic} is a high resolution neutrino cross section experiment in the NuMI beam line at 
Fermilab, upstream of the MINOS near detector.  
The goal of the experiment is to measure exclusive and inclusive neutrino cross 
sections in the energy range of 1-20 GeV on several nuclei with greatly improved 
precision over previous experiments.  MINERvA will be able to address several issues
raised above.  It will span the energy range between MiniBooNE and NOMAD, allowing
a study of the extracted $M_A$ as a function of energy and on different nuclear targets,
which will allow a study of the possible nuclear effects in the extraction of $M_A$.
It will also have good sensitivity to CC coherent production over the range of neutrino
energies where models appear to fail.

\begin{figure}[ht]
 \includegraphics[width=65mm]{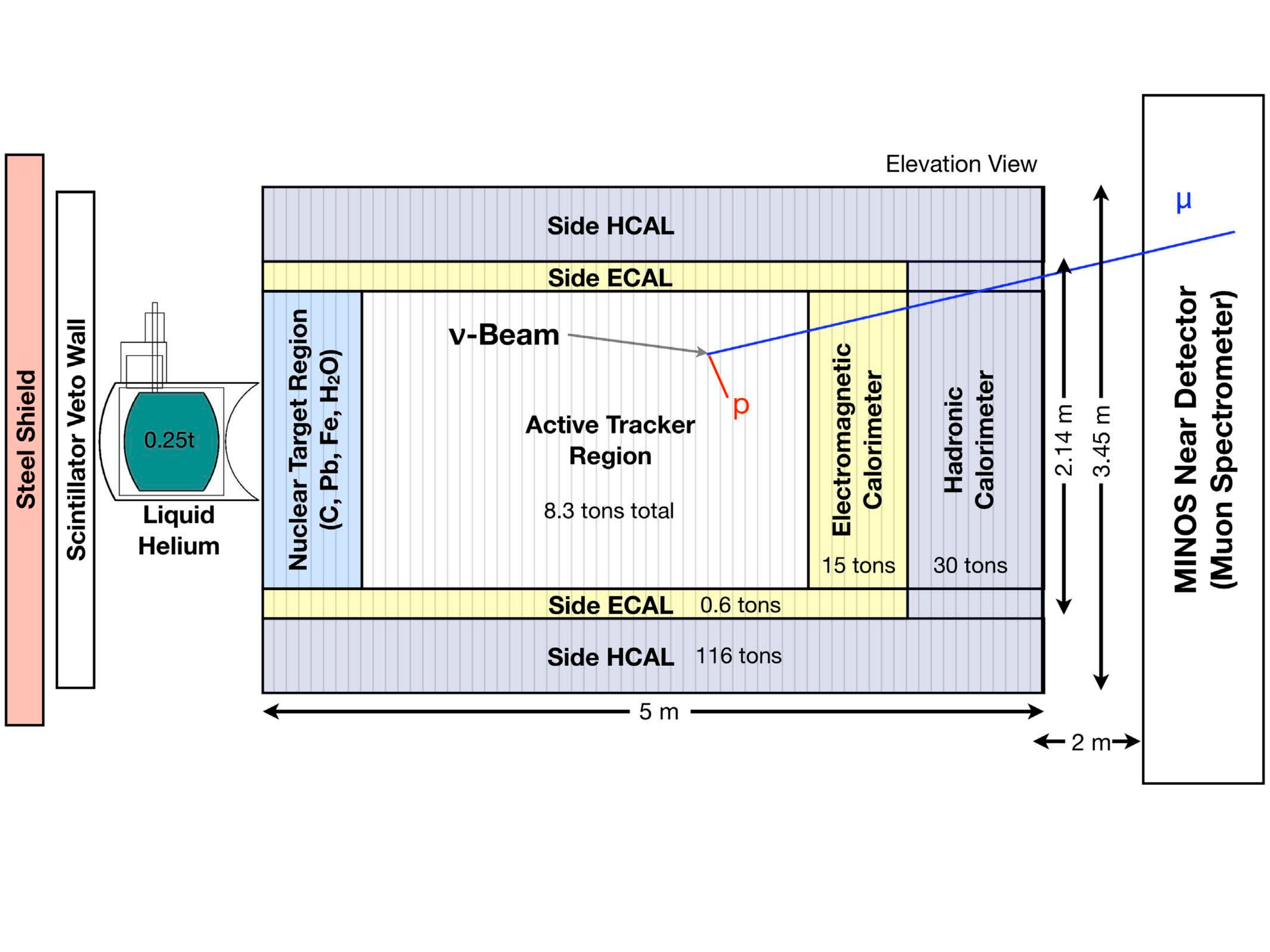}%
 \caption{Schematic side view of the MINERvA detector.}
\label{minerva-side}
 \end{figure}

A schematic of the MINERvA detector is 
shown in Fig.~\ref{minerva-side}.  The detector consists of five main regions: the fully active 
central detector, the upstream nuclear targets, a downstream electromagnetic and 
hadron calorimeter, and a surrounding electromagnetic and hadron calorimeter.  

The central detector serves as both the primary target and the tracking detector.  
It consists of planes of triangular plastic scintillator strips arranged in three orientations.  
Each strip's triangular cross section
 is 1.7 cm high and 3.3 cm wide, and is read out via a wavelength shifting fiber.  The strips
range from about 1 m to 2 m in length.
Light sharing between the strips gives a position resolution of approximately 3 mm.  
The light yield is approximately 5.0 photo-electrons/MeV, giving about 13.5 photo-electrons/plane 
for minimum ionizing muons.

The downstream electromagnetic calorimeter consists of alternating planes of 2 mm thick Pb 
and scintillator planes of the same hexagonal shape as in the central detector.  The hadron 
calorimeter is similar, with 2.5 cm planes of steel instead of Pb.  The side 
electromagnetic calorimeter consists of 2 mm thick Pb plates between tracking 
plane in the outer region of the central detector.  The side hadron calorimeter consists of 
planes of steel with scintillator strips embedded.

\begin{figure}[ht]
 \includegraphics[width=65mm]{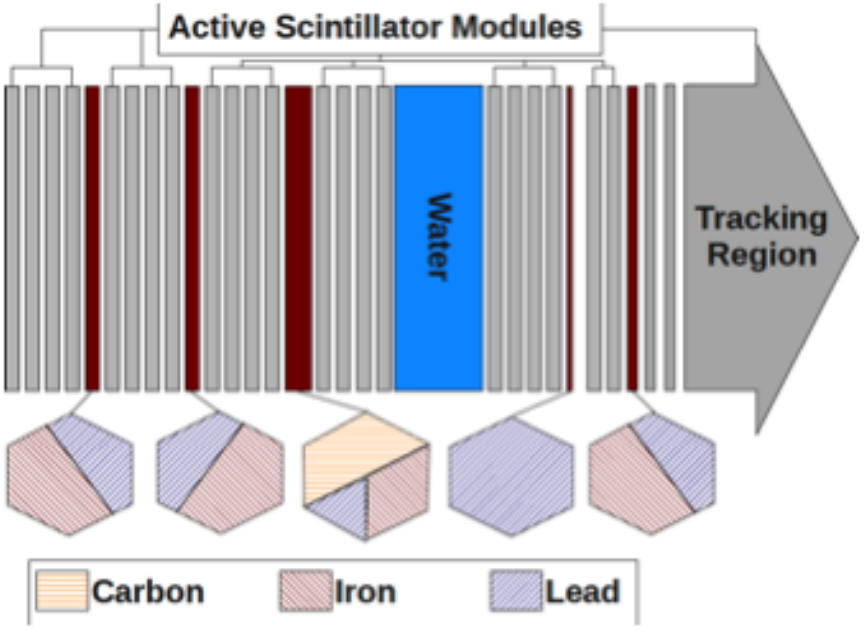}%
 \caption{Nuclear target region of the MINERvA detector.}
\label{minerva-nuc}
 \end{figure}

Upstream of the central detector are planes of passive targets, with two planes 2.5 cm 
thick of mixed Fe/Pb, one plane with 2.5 cm thick Fe/Pb and 7.5 cm C, a solid plane 
of Pb 0.80 cm thick, and a mixed plane of Fe/Pb 1.30 cm thick.  The mixed Fe/Pb planes 
are split with part of the plane being iron and part of the plane being lead, such that 
the total mass is approximately equal.  Tracking planes are placed between each plane of 
passive targets. Planned for fall 2011 are a 15 cm water target to be installed downstream 
of the Fe/Pb/C target and a tank of liquid 4He about 1 m in diameter to be installed 
upstream of the main detector.  Figure~\ref{minerva-nuc} shows a schematic of the nuclear target region 
along with the placement of carbon, lead, and iron in the various targets.

Charged current events originating in the central detector are fully contained, except for the muon, 
for neutrino energies of less than about 10 GeV.  The MINOS detector gives both muon energy and 
charge for forward going muons.  The threshold energy for a muon to exit MINERvA and be 
tracked into MINOS is about 2 GeV.  There is good angular acceptance for muons with 
scattering angles below 10 degrees, with acceptance dropping to zero for scatters greater 
than about 20 degrees.   For particles stopping in MINERvA particle identification can be 
determined from the dE/dX, but there is no charge determination.

MINERvA began operations
 with about half of the detector installed in November, 2009, and took data with the 
anti-neutrino beam until March, 2010.  Installation of the full detector was completed 
in March, 2010.  Running since that time has been divided between the neutrino mode, 
anti-neutrino mode, along with a few special settings intended to help determine the 
neutrino flux. The analysis to date has concentrated on those events which have a muon 
identified in MINOS.  Two of the first analyses are of the A dependence 
of the inclusive cross section and of the quasi-elastic anti-neutrino cross section.

\begin{figure}[ht]
 \includegraphics[width=65mm]{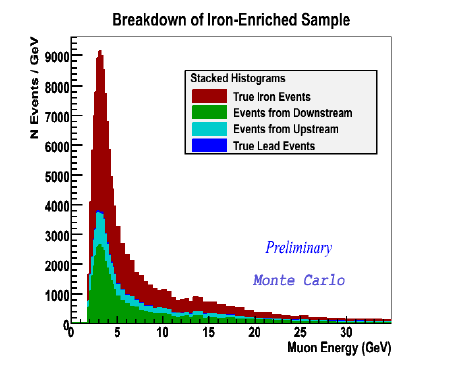}%
 \caption{A Monte Carlo estimate of the relative number of events tracked
to the iron target vs. the actual vertex location.}
\label{fe-bkgd}
 \end{figure}

MINERvA will be able to determine the nuclear dependence of the inclusive cross section, 
and in particular the relative ratio of lead, iron, and carbon. Although the statistical
precision wll be quite good, better than 1\%, the measurement requires a precise 
determination of the actual interaction nucleus in a given interaction.  The transverse 
resolution of the detector is quite good, and Monte Carlo simulations indicate that fewer 
than 0.1\% of vertices will be tracked to the wrong target.  The longitudinal resolution 
is less precise and depends on the number of final state particles and how far they travel.  
Because there is no tracking in the Fe/Pb targets, it is not possible to tell if events with 
a single muon in the final state originated in Pb/Fe target or in the downstream tracking 
detectors.  Monte Carlo studies of the background do indicate that it is substantial, as shown
in Fig.~\ref{fe-bkgd}. 

\begin{figure}[ht]
 \includegraphics[width=65mm]{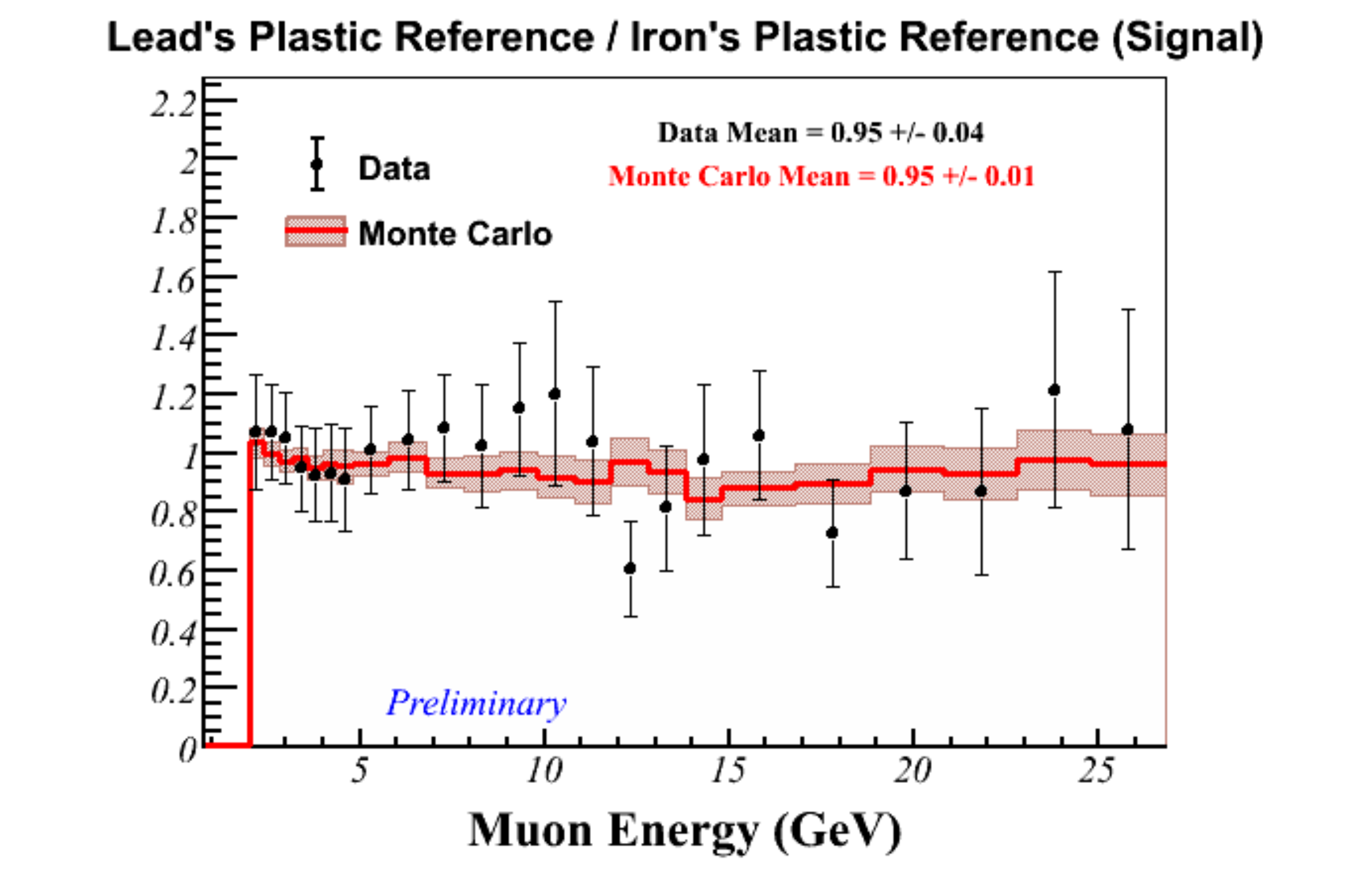}%
 \caption{The ratio of the iron-like to lead-like regions of the tracking detectors
as a function of muon energy, compared to Monte Carlo.}
\label{mu-ratio}
 \end{figure}

A measurement of the cross section in iron requires subtraction of the background from
scintillator.  The cross section on scintillator can be determined from the tracking
detectors, and is currently underway.  In addition to the background subtraction,
an acceptance correction must be made.  The acceptance of muons from the Pb and Fe targets 
is not identical, so a precise measurement of the Fe/Pb ratio requires a precise 
acceptance correction, which is done with the Monte Carlo simulation.
 The Monte Carlo has been checked by comparing data from regions of scintillator with the 
same areas as the Fe/Pb just downstream of the target, and comparing with Monte Carlo estimates.  
There is good agreement between data and Monte Carlo,  as shown in Fig.~\ref{mu-ratio}.  
Analysis of the data to determine the Pb/Fe inclusive ratio is still underway.

\begin{figure}[ht]
 \includegraphics[width=65mm]{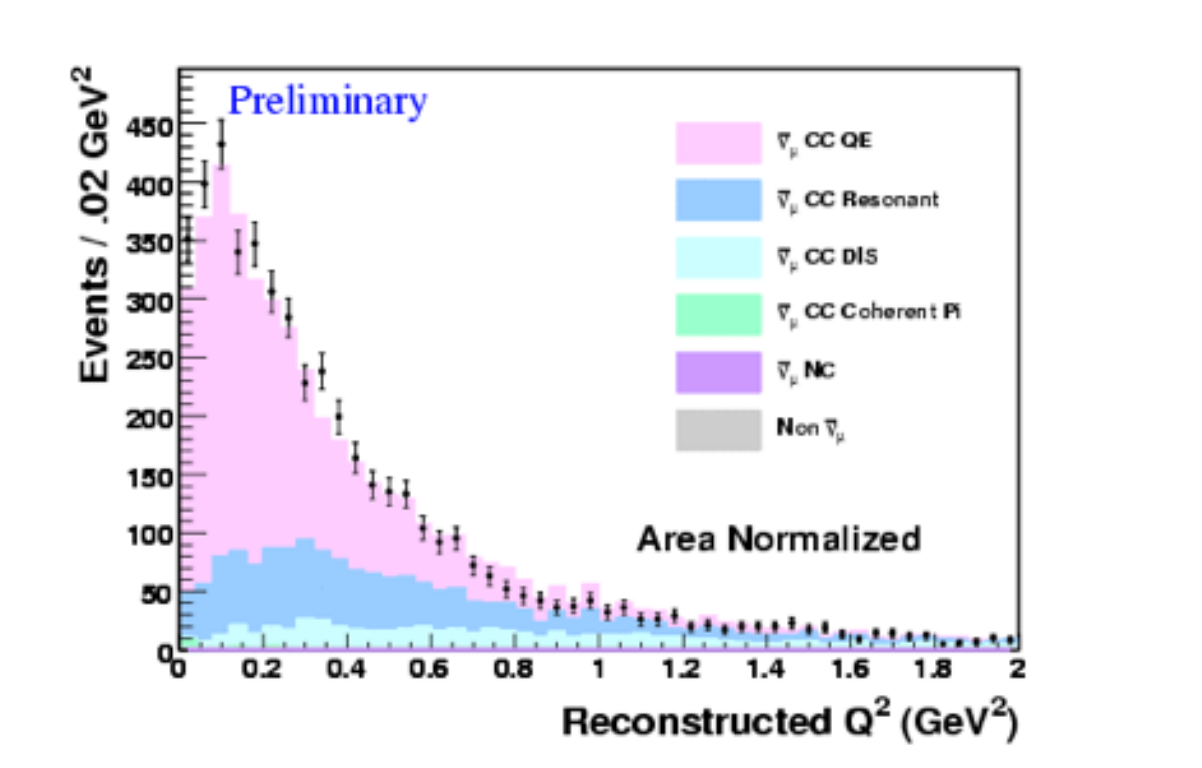}%
 \caption{Anti-neutrino quasi-elastic cross section compared to a Monte Carlo estimate,
summed over all neutrino energies.  The Monte Carlo is area normalized to the data.}
\label{qsq-alle}
 \end{figure}

The anti-neutrino quasi-elastic scattering was measured in the first running period with the 
partial detector.  Figure~\ref{qsq-alle} shows the $Q^2$ distribution for all energies combined
compared to a Monte Carlo simulation of the components of the measured cross section, with
the Monte Carlo results normalized to the data.  The 
preliminary data are in general agreement with an axial mass of 1.05.   
However, studies of the absolute neutrino flux, as well 
as further analysis of the scattering data, are still underway.  The statistics of the
data shown are only about one-eighth of the total
anticipated anti-neutron data, and 1/30 of the total neutrino data anticipated.  The precision,
especially at higher $Q^2$, will make it possible to compare the shape of the axial form factor to the
dipole form for the first time.

\subsection{T2K}

\begin{figure}[ht]
 \includegraphics[width=65mm]{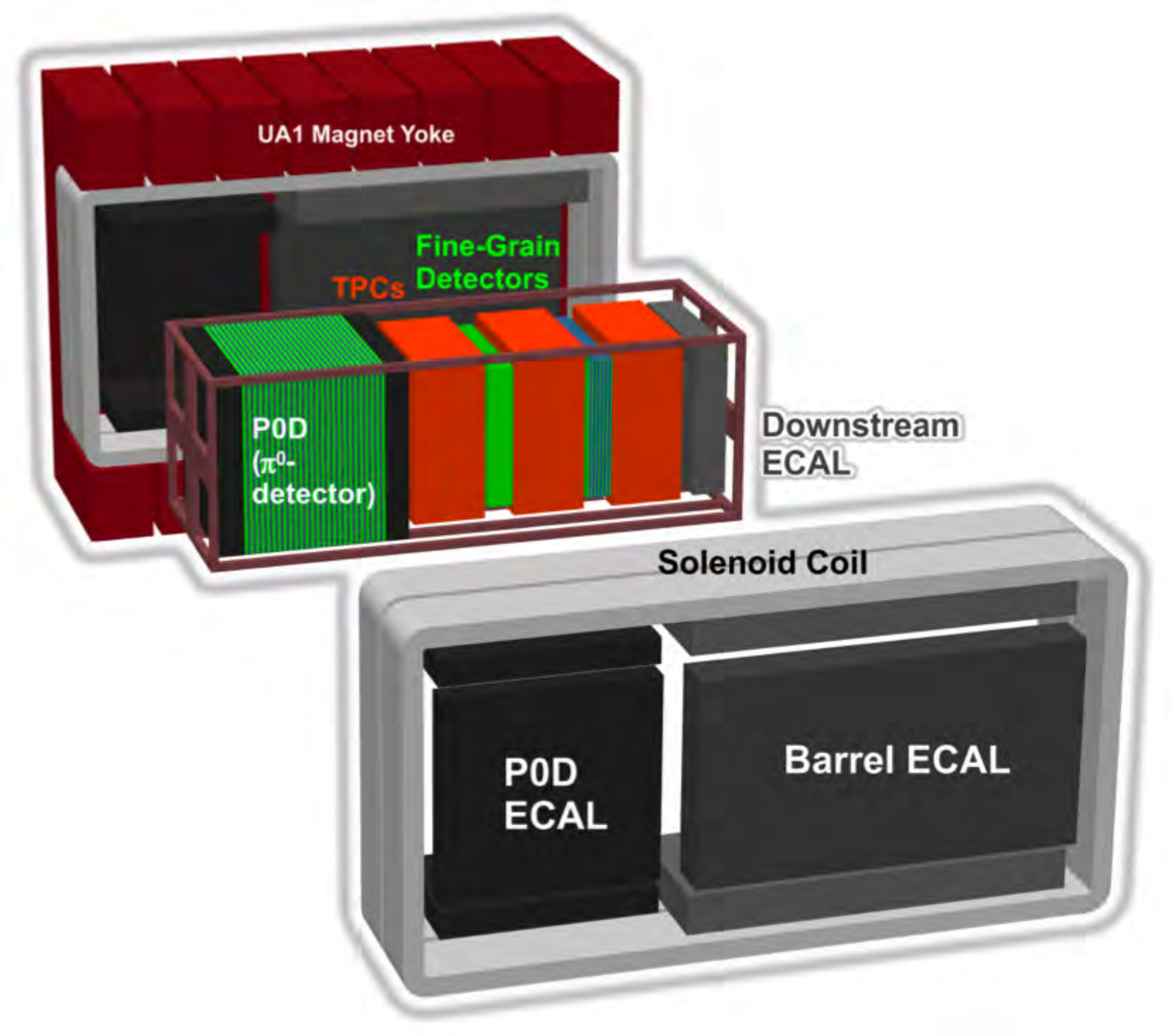}%
 \caption{T2K near detector.}
\label{t2kND}
 \end{figure}

\begin{figure}[ht]
 \includegraphics[width=65mm]{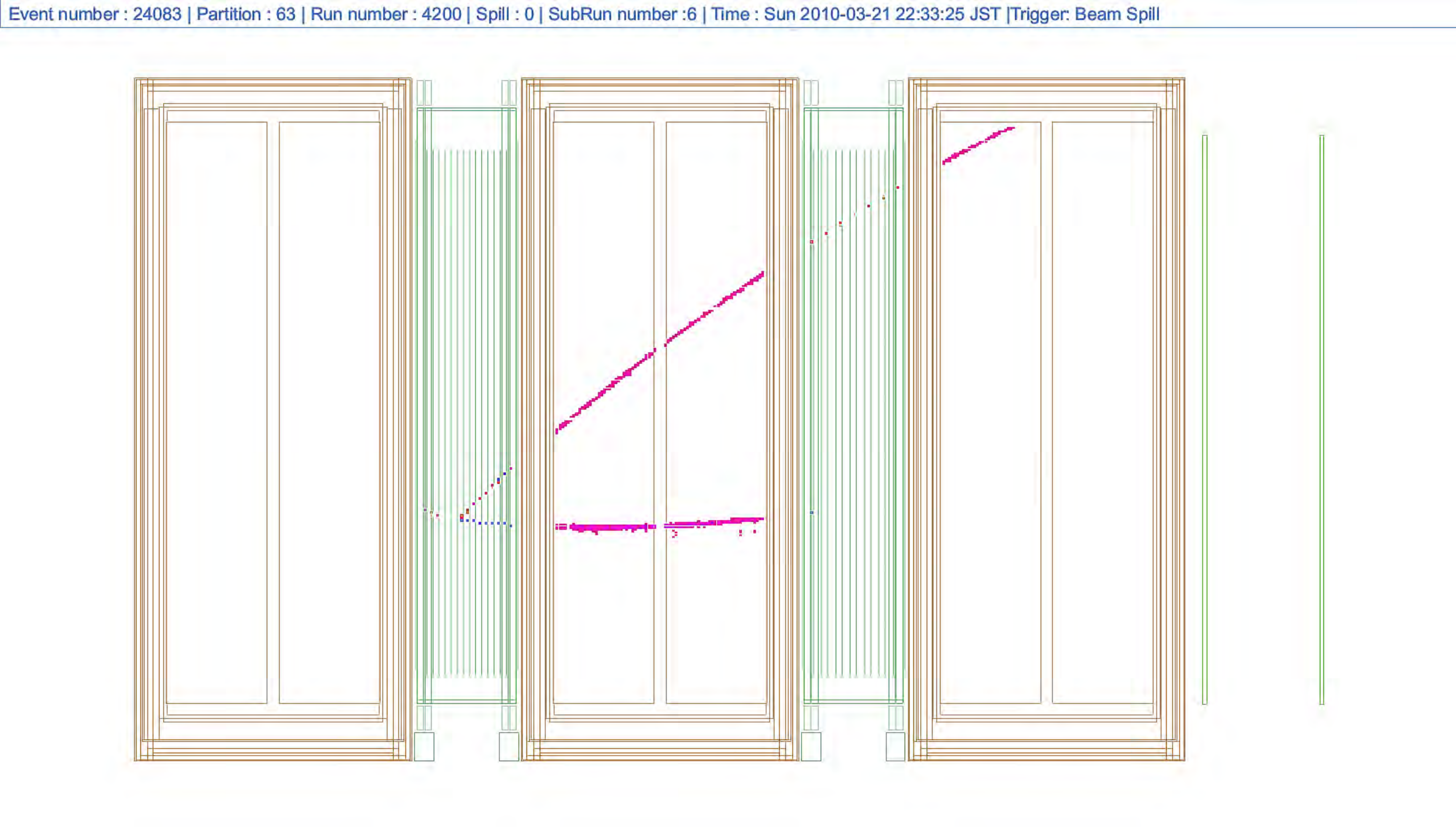}%
 \caption{T2K two track event in ND280 detector.}
\label{t2kevent}
 \end{figure}

A second experiment which will substantially improve our knowledge of low energy interaction cross
sections is T2K at JPARC.  Although primarily designed to measure the $\theta_{13}$
oscillation parameter, T2K also has a near detector, called the ND280 detector.  A schematic of
the detector is shown in Fig.~\ref{t2kND}.  ND280 includes an upstream $\pi^0$ dectector (P0D),
consisting of interleaved scintillator and water targets, designed to measure $\pi^0$
production in water.  Downstream of the P0D are 
fine grained scintillator detectors, consisting of 1 cm square scintillator bars which
are used as active neutrino targets, interleaved with time projection
chambers (TPCs) for particle identification.  The inner detectors are 
surrounded by electromagnetic calorimetery and the entire assembly is contained inside the former
CERN UA1 magnet, which provides a nominal 0.2 T field.  A sample event is shown in Fig.~\ref{t2kevent}.

The ND280 detector is 2.5 degrees off axis of the neutrino beam, giving a peak energy of
about 600 MeV with a FWHM of about 400 MeV.  This relatively low energy and relatively
narrow beam, coupled with the very good $\pi^0$ detection, excellent tracking,
charged particle identification and good energy measurements provided by the FGD will allow
T2K to study neutrino interactions carbon and oxygen for sub-GeV neutrino
energy with unprecedented precision.  

T2K began data taking in early 2010 and continued until neutrino production stopped following
the Fukushima earthquake in March 2011.  Data taking should begin again in early 2012.

\section{SUMMARY}

A new generation of high intensity neutrino beams and fine grained detectors have made
possible the use of the neutrino as a probe of nucleon structure.  As the precision
of measurements has increased, significant discrepancies between the data from different
experiments as well as comparisons between data and theoretical predictions.  The understanding
of neutrino interactions in nuclei has become increasingly important, and will be addressed
by new high precision experiments in the coming years.

\bigskip 
\begin{acknowledgments}

This work supported by the US National Science Foundation and the US Dept. of Energy.
I would like to thank the MINERvA collaboration, the T2K collaboration, B. Berger, 
M. Wascko, R. Petti, and G. Zeller for assistance with this presentation.

\end{acknowledgments}

\bigskip 
\bibliography{basename of .bib file}

\begin{thebibliography}{99} 
%
%
\bibitem{boone} BooNE website: http://www-boone.fnal.gov/
\bibitem{jparc} JPARC website: http://j-parc.jp/index-e.html
\bibitem{numi} NuMI website: http://www-numi.fnal.gov/
\bibitem{cngs} CNGS website: http://proj-cngs.web.cern.ch/proj-cngs/
\bibitem{Wilk2010} M.~Wilking, Physics in Collision, Proceedings, 30th International Conference,
PIC 2010, Karlsruhe, Germany, J. Komaragiri, T. Muller, J. Wagner-Kuhr (editors) 190,  2011.
\bibitem{Zeller} G. Zeller, arXiv:hep-ex/0312061 (2003).
\bibitem{NOMAD} J. Altegoer, {\it et al.}, Nucl. Inst. Meth. {\bf A404}, 96 (1998).
\bibitem{Lyub2009} V. Lyubushkin {\it et al.}, Euro. Phys. J. C {\bf 63} 355 (2009).
\bibitem{miniboone} A.A. Aguilar-Arevalo, Phys. Rev. D {\bf 81}, 092005 (2010).
\bibitem{sciboone} M.O. Wascko, Nucl. Phys. B Suppl. {\bf 00}, 1 (2011), J. Alcaraz
and J. Walding, arXiv:0909.564 [hep-ex] (2009).
\bibitem{ben} Figure courtesy B. Ziemer, UC Irvine.
\bibitem{gegm} See A.J.R. Puckett, {\it et al.}, Phys. Rev. Lett. {\bf 104},
242301 (2010), G. Ron, {\it et al.}, Phys. Rev. Lett. {\bf 99}, 202002 (2007), and
C.F. Perdrisat, V. Punjabi, M.Vanderhaeghen, Prog. Nucl.Part. Phys. {\bf 59},
694 (2007) and references therein.
\bibitem{arrington} J. Arrington, W. Melnitchouk, J.A. Tjon, Phys. Rev. C {\bf 76},
035205 (2007).
\bibitem{paolone} M. Paolone, {\it et al.}, Phys. Rev. Lett. {\bf 105}, 072001 (2010).
\bibitem{saito} K. Saito, K. Tsushima, and A.W. Thomas, Prog. Part. Nucl. Phys. {\bf 58}, 1 (2007).
\bibitem{carlson} J. Carlson, J. Jourdan, R. Schiavilla, and I. Sick, Phys. Rev. C {\bf 65},
024002 (2002).
\bibitem{bodek} A. Bodek, H.S. Budd, and E. Christy, arXiv:1106.0340[hep-ph], and Eur. Phys. J. C
{\bf 71}, 1726 (2011). 
\bibitem{nova} Website: http://www-nova.fnal.gov/
\bibitem{MINOS} See http://www-numi.fnal.gov/ for a description of the MINOS experiment.
\bibitem{models}S.S. Gershtein, Yu.Ya.Komachenko, M.Y. Khlopov, Sov. J. Nucl. Phys. {\bf 32}, 861 (1980);
Yu.Ya. Komachenko and M.Yu. Khlopov, Sov. J. Nucl. Phys. {\bf 45}, 295 (1987); B.Z. Kopeliovich and P. Marage,
Int. J. Mod. Phys. {\bf 8} 1513 (1993); 
S.K. Singh, M.S. Althar, S. Ahmad, Phys. Rev. Lett. {\bf 96}, 241801 (2006);
E.A. Paschos, A. Kartavtsev, and G.J. Gounaris, Phys. Rev. D {\bf 74}, 
054007 (2009); C. Berger
and L.M. Sehgal, Phys. Rev. D {\bf 79}, 053003 (2009);  M. Martini, M. Ericson, G. Chanfray, and
J. Marteau, Phys. Rev. C {\bf 80}, 065501 (2009);
E. Hernandez, J. Nieves, and M.J.V. Vascas, Phys. Rev. D {\bf 80}, 013003 (2009);
S.X. Nakamura, {\it et al.}, Phys. Rev. C {\bf 81}, 035502 (2010).
\bibitem{rein} D. Rein and L.M. Sehgal, Ann. Phys. {\bf 133}, 79 (1981), D. Rein and L.M. Sehgal,
Nucl. Phys. {\bf B223}, 29 (1983), D. Rein and L.M. Sehgal,
Phys. Lett. B {\bf 657}, 207 (2007).
\bibitem{alvarez} L. Alvarez-Ruso, NuInt11 proceedings (to be published) and www.nuint11.in.
\bibitem{boyd}S. Boyd, {\it et al.} in 
 AIP Conf. Proceedings 1189, F. Sanchez, M. Sorel, L. Alvarez-Ruso, A. Cervera, M. Vicente-Vacas,
eds., (Melville, NY) 60.
\bibitem{MINOS-coh} D. D. Cherdack, lss.fnal.gov/archive/thesis/fermilab-thesis-2010-54.pdf and D.D. Cherdack,
NuInt11 presentations, http://nuint11.in/
\bibitem{k2k-coh} M. Hasegawa, {\it et al.}, Phys. Rev. Lett. {\bf 95}, 252301 (2005).
\bibitem{scib-coh} Y. Kuitmoto, {\it et al.}, Phys. Rev. D {\bf 81}, 111102(R) (2010).
\bibitem{Hern} E. Hernandez, J. Nieves, M.Valverde, Phys. Rev. D {\bf 82}, 077303 (2010).
\bibitem{min-coll}MINERvA collaboration members listed at http://minerva.fnal.gov/.
\bibitem{panic} R.D. Ransome, PANIC 2011 proceedings, to be published.
\bibitem{t2k-berg} B. Berger, PANIC 2011 proceedings, to be published, and V. Paolone and S. Dytman, 
T2K-PROC-008, www.t2k.org/docs/proc/index\_html.

%
\end{thebibliography}

\end{document}